# Wideband Mid Infrared Absorber using surface Doped Black Silicon

S. Sarkar[1,2], E. Nefzaoui[1, a)], G. Hamaoui[1], F. Marty[1], P. Basset[1], T. Bourouina[1]

[1)] *ESYCOM, Univ Gustave Eiffel, CNRS, F-77454 Marne-la-Vallée, France*
[2)] *Current Affiliation: Dept. of Physics and Materials Science, University of Luxembourg, L-1511 Luxembourg, Luxembourg*

Black silicon (BSi) is a synthetic nanomaterial with high aspect ratio nano protrusions inducing several interesting properties such as a very large absorptivity of incident radiation. We have recently shown that heavily doping the BSi in volume enables to significantly enhance its mid infrared absorptivity and tune its spectral range of interest up to 20 µm. In the present letter, we explore the effect of surface doping on BSi radiative properties and it absorptance, in particular since surface doping enables reaching even larger dopant concentrations than volume doping but at more limited penetration depths. We considered 12 different wafers of BSi, fabricated with cryogenic plasma etching on n and p-type silicon wafers and doped using ion-implantation with different dopant types, dosages and ion beam energies leading to different dopant concentrations and profiles. The different wafers radiative properties, reflectance, transmittance and absorptance, are experimentally measured using Fourier transform infrared spectroscopy. We show that doping an n-type BSi wafer with Phosphorous with a dose of $10^{17}$ atm/cm$^2$ and an energy of 100 keV increases its absorptivity up to of 98% in the spectral range of 1-5 µm. We propose a simple phenomenological explanation of the observed results based on the dopant concentration profiles and the corresponding incident radiation penetration depth. Obtained results provide simple design rules and pave the way for using ion-implanted BSi for various applications such as solar energy harvesting, thermo-photovoltaics and infrared radiation sensing where both high absorptance and variable dopant concentration profiles are required.

The enhancement of waste and renewable energy use by increasing the efficiency of energy conversion devices is a particularly active field of research today, in order to cope with the rising societies' energy needs and to tend towards a carbon-free society. For this purpose, metamaterials are widely considered as promising candidates since they offer the possibility of designing new materials with properties that cannot be found in natural materials and optimized for specific applications. In addition, 90% of the energy production worldwide involves the manipulation of heat. Consequently, meta-materials for thermal energy conversion and management has been a hot topic in the past two decades because of their huge potential in many applications such as thermoelectric, photovoltaic and thermo-photovoltaic energy conversion, thermal management of electronic devices, thermal rectification, large scale thermal management using radiative cooling, etc. Amongst the explored meta-materials, silicon-based materials have received a particular attention, taking benefit of the wide use of silicon in the microelectronic industry, which provides several advantages such as the ease of fabrication, system scale integration, and affordable scalability. In this context, silicon has been used for the design and fabrication of phononic[1] and photonic[2] crystals, resonating cavities[3], 1D[4,5] and 2D[6] surface gratings as well as porous metamaterials[7]. In this letter, we focus on a peculiar silicon metamaterial commonly called Black Silicon (BSi), which refers to silicon surfaces covered by a layer of nano- or fine micro-structures that effectively suppresses reflection while increasing light absorption[8]. As a result, the silicon wafers seem black rather than the silver-grey typical of flat silicon (FSi) wafers. For this purpose, BSi has initially been identified as a good candidate to enhance light absorption in the visible range and consequently to enhance the performance of solar energy harvesting devices[9,10]. The peculiar surface features of BSi provide it with additional appealing qualities such as low reflectance, large and chemically active surface area, super hydrophobicity, and high luminescence efficiency. For these reasons, it has been recently used for multiple applications ranging from solar cells[10] and light-emitting devices[11] to antibacterial coatings[12] and gas sensors[8].

To go beyond the visible range applications of BSi, various methodologies have been explored to enhance BSi radiative properties, such as gold nano-particle assisted BSi substrates for mass spectrometry imaging,[13,14] optical tweezers[15], broadband absorption,[16] silver-coated BSi for enhanced Raman scattering,[17,18] BSi on stainless steel foil for broadband absorption,[19] etc. We have recently studied the impacts of heavy volume doping of silicon as well as its morphological implications on BSi surface features and radiative properties.[20] We have shown, without any need for additional surface functionalization, that heavily volume doped BSi exhibits a larger absorptivity and emissivity over a broader and tunable spectral range.[20] Since doping is a governing parameter in BSi morphology and radiative properties, reaching the highest doping levels can be useful to further enhance the BSi absorptivity and for tunability

[a)] Email: elyes.nefzaoui@univ-eiffel.fr

purposes. However, the volume doping levels achieved during silicon growth are limited due to the solid solubility limits of a given dopant in silicon. One way to overcome this limit and reach higher levels of doping is to use surface doping. Ion implantation has emerged as the leading technique.[21] Ion implantation is a material surface modification method in which ions of one material are implanted into another solid material, resulting in a change in the physical and chemical properties of the materials' surfaces,[21–24] without altering their bulk material properties. Ion implantation is a 'top-down' approach to semiconductor doping, widely used because of the ease of the ion energy selection and surface mask application for in the depth and spatial distribution of doping substances respectively.[24] However, this fabrication method suffers a major drawback since the obtained doping concentration rapidly decreases with increasing depth inside the sample.[25]

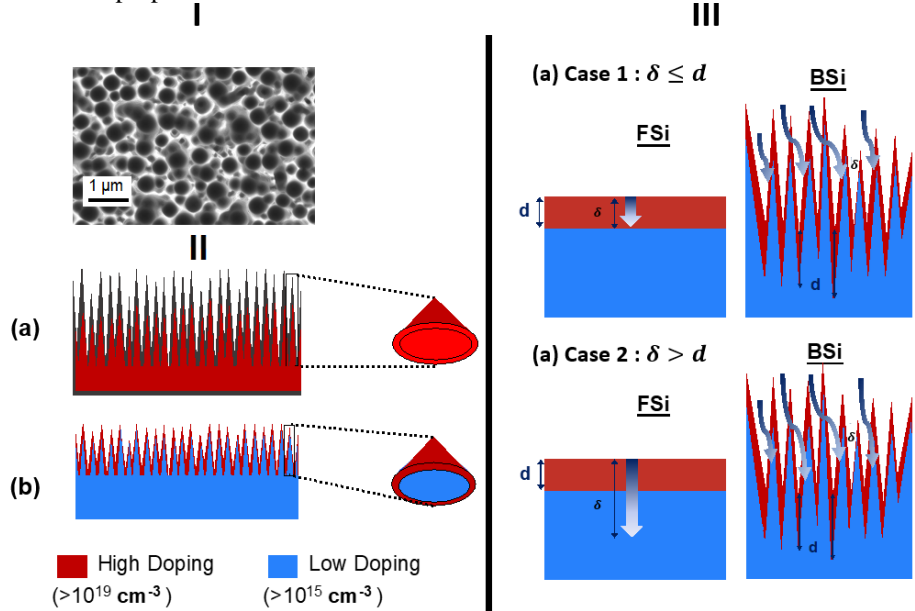

FIG. 1 (I): Top view SEM image of a fabricated sample of BSi; (II) Schematic of (a) volume doped and (b) surface doped BSi showing their differences; (III) illustration of the threshold depth ($d$) and skin depth ($\delta$) for flat Si and BSi for the two cases: (a) $\delta \leq d$ (b) $\delta > d$.

In this letter, we report on an investigation on the radiative properties. Doping is achieved through ion-implantation at different doses and energy levels over a raw black silicon surface, typical of Fig 1(I). With surface doping, doping concentrations of $10^{20}$ at/cm$^3$ and beyond can be reached which expands the scope of volume doping in commercially available silicon substrates. This was the first motivation to fabricate and investigate surface doping on BSi. A subsidiary question is whether surface doping is sufficient or not to achieve the ultra-broadband absorption we recently reported in heavily volume doped BSi. Indeed, surface doping via ion-implantation on BSi enables reaching high concentration levels only at the substrate surface within a thickness not exceeding 1 µm and even lower. We schematically show in Fig 1(II) the difference in doping profiles, in particular with respect to BSi surface features depth, between heavily volume doped BSi and heavily surface doped BSi. BSi wafers were fabricated starting from n-type and p-type silicon with a resistivity of 1-20 ohm.cm using cryogenic deep reactive ion etching process developed in our prior report[26]. The process is carried out in a A601E Alcatel etch tool configured for cryogenic plasma silicon etching at a temperature of -110°C. One should note that the ion implantation at the surface of BSi, shown in Fig 1(II.b), is achieved after the fabrication of raw BSi by nanostructuration of a low doping wafer. The resulting surface keeps its initial topography unaffected by ion implantation, whose result is a conformal doping of the nanostructured silicon. The ion implantation process is governed by the following five parameters: dopant species, ion beam energy, implantation dose, and tilt and twist angles. We have considered 12 BSi silicon wafers which have been subjected to various degrees of ion-implantation based on three principal parameters namely the type of dopant, ion beam energy (KeV), and implantation dosage (at/cm$^2$).[27–30] The ion implantation has been done by Ion-Beam Services (IBS) company, France.

All the aforementioned parameters have been considered to comprehend their impacts on BSi doping profiles radiative optical properties and to enable a tuning route for obtaining a suitable absorber for a specific application. The doping profiles across the depth of the ion-implanted samples have been numerically calculated as a function of the ion-implantation parameters.

On the basis of the obtained dopant concentration profiles, we have calculated the *threshold depth d* defined as the depth after which the ion implanted BSi fails to retain the highest concentration level of the order of $10^{20}$ cm$^{-3}$. Since we are primarily interested in values of doping larger than $10^{19}$

cm$^{-3}$, this threshold depth is an important criterion in establishing optimum ion-implantation parameters to maximize radiation absorption at the sample surface in a depth lower than 1 µm. Thereafter, the *skin depth,* defined as the depth at which the intensity of the radiation inside the material falls to $1/e$ of its original value at the surface, is calculated for each sample. It can be expressed as:

$$\delta = \sqrt{\frac{2\rho}{\omega\mu}}\sqrt{\sqrt{1+(\rho\omega\varepsilon)^2}+\rho\omega\varepsilon} \quad (1)$$

where $\rho$ is the resistivity of the material; $\omega$ the angular frequency $= 2\pi f$, $f$ the frequency; $\mu$ the permeability of the material, $\mu = \mu_r\mu_0$ along $\mu_r$ the relative magnetic permeability of the material, and $\mu_0$ the permeability of free space; $\varepsilon = \varepsilon_r\varepsilon_0$ the permittivity of the material with $\varepsilon_r$ the relative permittivity of the material, and $\varepsilon_0$ the permittivity of free space. We have calculated the skin depth for particular values of resistivity for p-type and n-type dopant namely, boron and phosphorous at the highest doping level of $10^{20}$ cm$^{-3}$, as this is the highest doping level achievable considering the solid solubility of the considered dopants in silicon. Then, we calculate the attenuation coefficient as $1 - e^{-2d/\delta}$, since the penetration depth of the electromagnetic waves in the material at normal incidence is proportional to the attenuation coefficient. Given the attenuation coefficient expression, we can have different situations where i) $d < \delta$; ii) $d = \delta$; and iii) $d > \delta$. The wavelength, $\lambda_c$, at which the condition of $d = \delta$ is fulfilled, gives us a critical value beyond which the absorption at the surface level will be reduced and can be used to tune the absorber spectral selectivity. These quantities are illustrated in Fig 1(III) and enable us not only to explain how light is absorbed in ion-implanted BSi, but also to maximize the absorptivity using the optimum values of implantation dosage and ion-beam energy in a particular range of wavelengths. Bearing these quantities in mind, we present and discuss the results obtained with various BSi ion-implanted samples in the following paragraphs.

Square samples of 1 cm² have been diced upon completion of the ion-implantation. The radiative properties, the transmittance T and reflectance R in particular, where then measured on the 1cm² samples by Fourier Transform Infrared (FTIR) Spectroscopy using a Perkin Elmer Spectrum 3 spectrometer with a DTGS detector in the spectral range from 1.3 µm up to 26 µm at room temperature. The absorptivity A is then computed using the energy conservation principle as $A = 1 - R - T$.

We present and discuss in the following paragraphs the results obtained with various BSi ion-implanted samples.

We show in Fig. 2(I), the measured reflectance (a), transmittance (b), and absorptance (c) of ion-implanted BSi samples with of an energy of 100 KeV of the n-type wafer in the spectral range from 1 to 5 µm. The samples' properties are grouped in Table 1.

First, regarding the samples' radiative properties, we observe that the heavily-dosed samples of W1 record the lowest level of reflectance of 0.96% at 2.5 µm and this rises to 1.84% at 5 µm. The level of reflectance rises as ion-implantation dosage decreases. Similarly, for transmittance, samples with the highest dosage exhibit the lowest transmittance. Among all the samples of the n-type wafer, it is observable that only those with heavy doses of ion-implantation, i.e., larger than $10^{16}$ at/cm² have transmittance below 5%. Interestingly, for samples 1 and 2, till 5 µm, there is a noticeable peak of transmittance which reaches 1.5%, 4.5%, and 5.3% and then reduces to 0.023%, 0.123%, and 0.25%, respectively for the entire wavelength range from 1.3-5 µm. All other samples rise above 15% of transmittance and the highest value is reached by the n-type reference Si sample. When compared to volume-doped n-type BSi samples (VDBSi Sample), no ion-implanted sample records such low transmittance levels ~ 0.02%. Regarding absorptance, the highest level of 98.86-96.35% in the wavelength range 1.3-5 µm, is reached by sample 1 with the highest dose of ion-implantation. The reference sample understandably records the lowest absorptance.

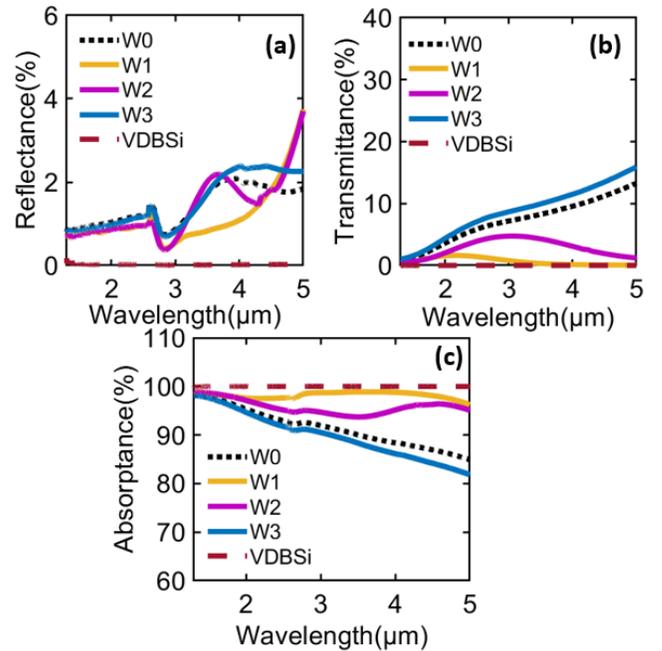

FIG. 2: Experimental results of reflectance (a), Transmittance (b) & Absorptance (c) of ion-implanted BSi samples of *n-type* wafers obtained with different ion-implantation dosages and the same energy (see Table 1) compared with a reference BSi sample (without ion implantation) and a heavily *n-type* volume doped BSi (VDBSi) sample in the wavelength range of 1-5 µm.

| Wafer ID | Energy (keV) | Dose (at/cm²) |
|---|---|---|
| W1 | 100 | $1.0 \times 10^{17}$ |
| W2 | 100 | $2.0 \times 10^{16}$ |
| W3 | 100 | $1.0 \times 10^{15}$ |
| W0 | Reference wafer (no ion-implantation) | |
| VDBSi | Volume doped N type | |

Table 1 : Ion-implantation parameters and wafer details of the studied samples.

One should note that for wavelengths larger than 5 µm, smaller absorptance is recorded for the different samples

(Supplementary Figure S1). We show in Fig. 3(a) the obtained dopant concentration profiles with ion-implantation simulations, phosphorous $P^+$ in this case, of the samples under consideration. We have seen in Fig 2, that W1 exhibits the lowest reflectance, lowest transmittance, and highest absorptance, followed by W2. From Fig 3(a), we extract the values of $d$ at a dopant concentration of $10^{20}$ cm$^{-3}$ and calculate the attenuation coefficient $1 - e^{-2d/\delta}$ shown in Fig 3(b). The dopant concentration for W3 does not reach the cut-off value of $10^{20}$ atm/cm$^3$, hence, does not appear in these figures. In Fig 3(b), we see that the attenuation constant is the highest for W1 followed by W2 and progressively drops with decreasing values of dosage and consequently threshold depths, $d$, indicating that below a dosage of $10^{16}$ atm/cm$^2$, a high attenuation coefficient cannot be achieved. For both W1 and W2, $d = \delta$ is fulfilled at a critical wavelengths, $\lambda_c$ = 3 µm and 4.5 µm, respectively. Going back to these samples absorptance at $\lambda_c$ = 3 (Fig. 2(c)), we observe that the absorptance falls below 98 % for both samples beyond $\lambda_c$. This indicates that this condition is indeed useful to determine the depth of high doping concentration required to obtain high absorption of incident radiation.

One can also conclude that ion-implantation with a dosage of $10^{17}$ at/cm$^2$ is sufficient to reach a complete absorption of light at the surface for BSi up to 5 µm. We can also note that this measure of the penetration depth is a key indicator than can be used to design and fabricate surface-doped BSi with enhanced absorptance in the spectral range of 1-5 µm. Surface doping alone can then be employed to considerably enhance silicon absorptivity, in wide mid infrared (MIR) range from 1 to 5 µm, even though it does not reach the performances of volume doped BSi in terms of absorptivity levels and the spectral range width.[20,31,32]

In addition, we also note (Supplementary Figure S2) that increasing levels of energy (keV) in the ion-implantation process increases the absorptance and reduces both reflectance and transmittance. Consequently, the largest absorptance is obtained with the largest ion-implantation energy, 100 keV in our case. This is consistent with the literature which reports that ion-beam energy is directly proportional to the depth of penetration.[33] However, the impact of energy is less conspicuous than the effect of dosage on the absorptance of BSi. Therefore, the dosage is the dominant factor to enhance the surface doped BSi absorptivity.

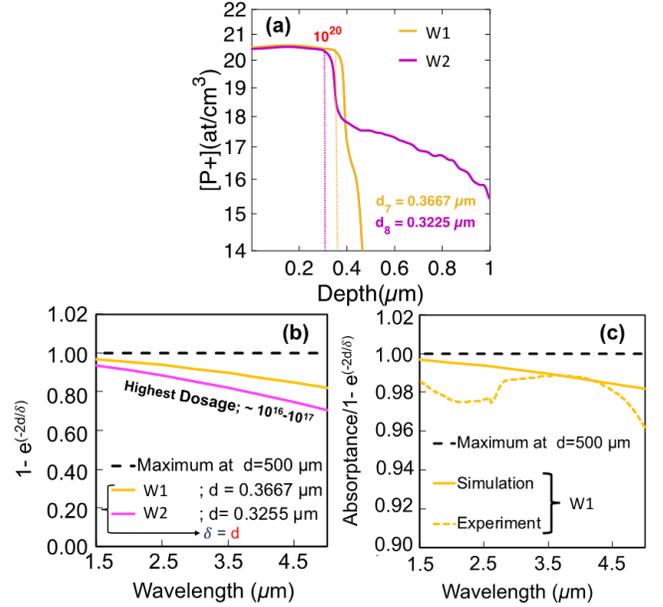

FIG. 3: (a) Simulation profiles of dopant (Phosphorous, $P^+$) concentration across depths of n-type BSi wafers after additional surface doping with ion-implantation, depicting the threshold depth $d$ at $10^{20}$ at/cm$^{-3}$ dopant concentration for each wafer; (b) Attenuation coefficient $(1 - e^{-2d/\delta})$ of W1 and W2, taking their corresponding values of $d$ and $\delta$. W3 is not shown as wafers with the highest dosage reach $1 - e^{-2d/\delta} > 0.90$ ; (c) Absorptance and attenuation coefficient for W1.

At last, we analyze the effect of the used dopant and the type of wafer. Two wafers of n-type and p-type (Samples 9 and 4) were implanted with opposite dopants i.e., n-type implanted with p-type dopant i.e., boron and p-type implanted with n-type dopant i.e., phosphorous. We observe (Supplementary Figure S3), that p-type wafer implanted with n-type dopant or has lower absorptance than n-type wafer implanted with p-type dopant. Wafers 4 and 9 have absorptance levels dropping to 85% and 65% respectively at 5 µm. The radiative properties i.e., reflectance, transmittance, and absorptance of n and p-type wafers at benchmark wavelengths of 5, 8, 10, 13 µm have been noted in Table SI and SII in the Supplementary Materials. In any case, all of these different combination lead to lower performances, if we consider large absorptivity as a target, than those presented and discussed above, if we consider the absorptance level as our figure of merit.

To summarize, surface doping on BSi can be employed to achieve considerably enhanced MIR absorptivity of silicon in particular in the spectral range 1-5 µm. Even though it does not lead absorption levels and spectral ranges as high and wide than those of heavily volume doped BSi, it offers the flexibility of altering only the samples surface, while keeping low doping concentration in the volume. This capability is very important for numerous applications such as photodetectors and solar cells for instance. In the case of surface doping, the dosage of ion-implantation is the dominant parameter for obtaining large absorptivity. In the present work, the highest levels of absorptance 98.9 % is provided by wafers subjected to highest phosphorous doping dosage of the order of $10^{17}$ at/cm$^2$. Higher ion-implantation energy levels also provide highest levels of absorptance but the results are less sensitive to energy than to the dosage. Quantitative analysis with dopant concentrations profiles, and an evaluation of the respective skin depths for each sample according to their threshold depths ($d$) for the highest concentration, attests to the fact that highest dosage provides the highest threshold depth, $d$. Only for samples where the dosage is of the order of $10^{17}$ at/cm$^2$ and specifically in our case for n-type wafer doped with phosphorous, the condition $d = \delta$ is fulfilled. This explains the highest absorptances recorded by these samples within 1-5µm. This indication provides a simple design rule to maximize the absorptivity of surface doped BSi by ion-implantation.

ACKNOWLEDGMENTS

This work was supported by the I-SITE FUTURE Initiative (reference ANR-16-IDEX-0003) in the framework of the project NANO-4-WATER.

# Supplemental material for: "Wideband Mid Infrared Absorber using surface Doped Black Silicon"

S. Sarkar[a,2], E. Nefzaoui[1 a)], G. Hamaoui[1], F. Marty[1], P. Basset[1], T. Bourouina[1]

[1)] *ESYCOM, Univ Gustave Eiffel, CNRS, F-77454 Marne-la-Vallée, France*

[2)] *Current Affiliation: Dept. of Physics and Materials Science, University of Luxembourg, L-1511 Luxembourg, Luxembourg*

This document contains additional information to the manuscript **"Wideband Mid Infrared Absorber using surface Doped Black Silicon".** It provides supplementary details on the different samples covered in the present study, with different doping types, levels, and on their radiative properties. Only a subset of the samples considered as the most promising was included in the main manuscript.

---

[a)] Email: elyes.nefzaoui@univ-eiffel.fr

1. **Radiative properties of the main samples over the extended MIR spectral range covered in the present study (1-25 μm);** here data are given for the complementary range of (5-25 µm), while the data for 1-5 µm range are given in the main manuscript.

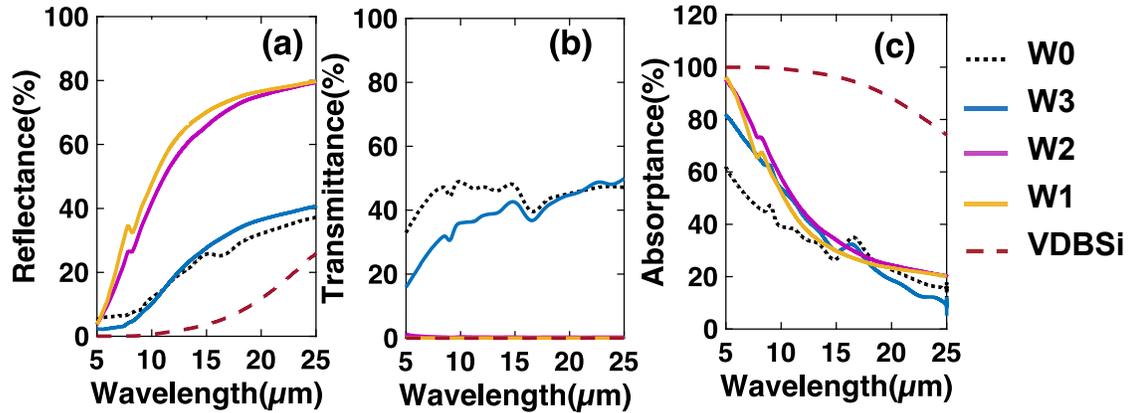

*Supplementary* **FIG S1:** Experimental results of reflectance (a), Transmittance (b) & Absorptance (c) of ion-implanted BSi samples of *n-type* wafer having different dosages but same energy compared with volume *n-type* doped BSi of doping level of $5 \times 10^{19}$ at/cm$^{-3}$ at a wavelength range of 5-25 µm is considered. When compared to Fig 2 of the main manuscript which shows 1-5 µm of wavelength range, the hierarchy is completely changed when it comes to 5-25 µm. S1 no longer provides the lowest reflectance, it now has the highest reflectance response along with S2. Consequently, the absorptance is low. Only transmittance responses remain unchanged for all wafers.

2. **Radiative properties of p-type BSi after p-type surface doping;** while the results in the main manuscript deal with n-type surface doping on n-type wafers, these additional results are obtained after p-type surface doping on p-type wafers

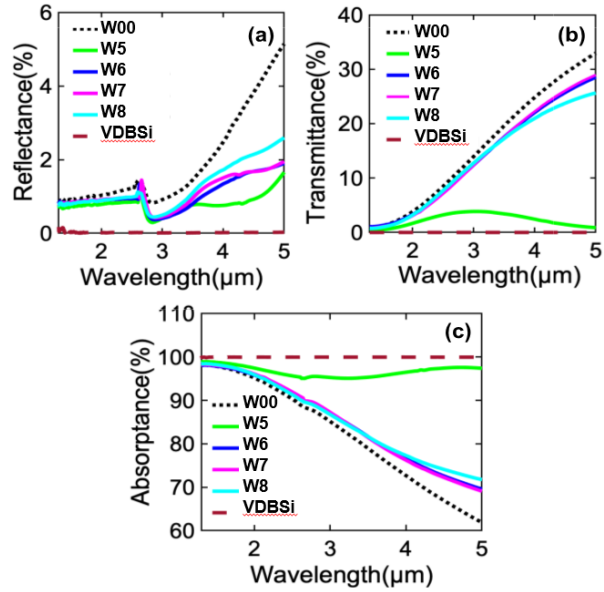

***Supplementary*** **FIG S2:** Experimental results of reflectance (a), Transmittance (b) and Absorptance (c) of ion-implanted BSi samples of *p-type* wafer having different dosages but same energy compared with *n-type* heavily volume doped BSi in a wavelength range of 1-5 µm.

| Wafer ID | Energy (keV) | Dose (at/cm²) |
|---|---|---|
| W5 | 100 | $1.0 \times 10^{17}$ |
| W6 | 100 | $2.0 \times 10^{15}$ |
| W7 | 50 | $2.0 \times 10^{15}$ |
| W8 | 30 | $2.0 \times 10^{15}$ |
| W00 | Reference p-type wafer (no ion-implantation) | |
| VDBSi | Volume doped n-type | |

***Supplementary Table 1 S2***: Ion-implantation conditions and wafer details of the p-type BSi samples (W5 to W8) and reference samples (W00 and VDBSi).

3. **Radiative properties of n-type and p-type BSi after p-type and n-type surface doping, respectively;** these additional results involve P-N junctions.

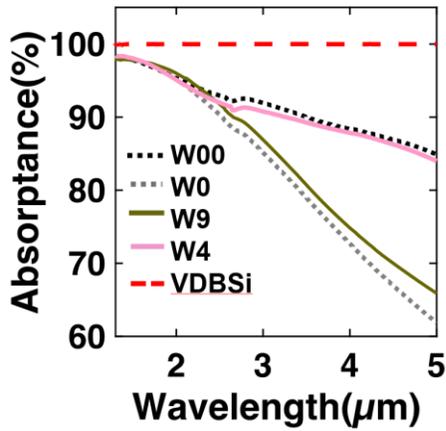

*Supplementary* **FIG S3:** (a) Experimental absorptance of ion-implanted BSi samples of an n-type wafer doped with p-type dopant (W4) and p-type wafer doped with n-type dopant (W9) with the same ion-beam energy of 70 keV and dosage of $10^{15}$ at/cm³ compared with heavily volume doped n-type BSi ($5.10^{19}$ cm⁻³) and the respective reference n-type (W0) and p-type (W00) BSi wafers in the wavelength range of 1-5 µm.

| Wafer ID | Wafer Type and resistivity | Type of dopant | Energy (keV) | Dose (at/cm²) |
|---|---|---|---|---|
| W4 | n-type 1-20 ohm.cm | B11 | 70 | $2.0 \times 10^{15}$ |
| W9 | p-type 1-20 ohm.cm | P31 | 70 | $1.0 \times 10^{17}$ |
| W0 | Reference n-type wafer (no ion-implantation) | | | |
| W00 | Reference p-type wafer (no ion-implantation) | | | |

*Supplementary Table 2 S3:* Ion-implantation conditions and wafer details of the samples considered in FIG S3.

# 4. Data table of reflectance, transmittance and absorptance for all samples at specific key wavelengths

*Supplementary* **TABLE 3 S4.** Reflectance, Transmittance and Absorptance of different n-type wafers after ion-implantation at specific wavelengths as a Figure of Merit for comparison purposes. The largest values of absorptance are indicated in red font.

| Wavelength (μm) | Wafer ID | Reflectance (%) | Transmittance (%) | Absorptance (%) |
|---|---|---|---|---|
| 1.3 | 0 | 0.83 | 0.89 | 98.3 |
| | 1 | 0.71 | 0.36 | 98.95 |
| | 2 | 0.70 | 0.5 | 98.8 |
| | 3 | 0.83 | 1.11 | 98.09 |
| | 4 | 0.87 | 0.84 | 98.31 |
| 2 | 0 | 1.04 | 3.67 | 95.33 |
| | 1 | 0.87 | 1.53 | 97.58 |
| | 2 | 0.86 | 2.04 | 97.09 |
| | 3 | 0.97 | 4.38 | 94.64 |
| | 4 | 0.96 | 5.45 | 95.09 |
| 5 | 0 | 1.84 | 13.3 | 84.85 |
| | 1 | 3.78 | 0.06 | 98.92 |
| | 2 | 3.69 | 1.202 | 95.09 |
| | 3 | 2.25 | 15.94 | 81.80 |
| | 4 | 2.81 | 13.2 | 83.98 |
| 8 | 0 | 8.84 | 25.56 | 65.59 |
| | 1 | 33.6 | 0.02 | 66.30 |
| | 2 | 26.53 | 0.34 | 73.12 |
| | 3 | 4.39 | 30.48 | 65.12 |
| | 4 | 8.09 | 20.29 | 71.61 |
| 10 | 0 | 15.09 | 30.19 | 54.72 |
| | 1 | 47.71 | 0.03 | 52.25 |
| | 2 | 41.82 | 0.36 | 57.81 |
| | 3 | 10.21 | 36.05 | 53.74 |
| | 4 | 13.25 | 22.49 | 64.26 |
| 13 | 0 | 22.41 | 32.53 | 45.06 |
| | 1 | 64.79 | 0.02 | 35.18 |
| | 2 | 59.37 | 0.29 | 40.33 |
| | 3 | 22.03 | 42.48 | 35.49 |
| | 4 | 23.49 | 20.96 | 55.55 |

*Supplementary* **TABLE 4 S4.** Reflectance, Transmittance and Absorptance of different p-type wafers after ion-implantation at specific wavelengths as a Figure of Merit for comparison purposes. The largest values of absorptance are indicated in red font.

| Wavelength (µm) | Wafer ID | Reflectance (%) | Transmittance (%) | Absorptance (%) |
|---|---|---|---|---|
| 1.3 | 00 | 0.88 | 0.89 | 98.22 |
|  | 5 | 0.72 | 0.375 | 98.90 |
|  | 6 | 0.82 | 1.043 | 98.13 |
|  | 7 | 0.81 | 0.87 | 98.31 |
|  | 8 | 0.81 | 0.73 | 98.44 |
|  | 9 | 1.44 | 0.72 | 97.83 |
| 2 | 00 | 1.02 | 2.67 | 96.3 |
|  | 5 | 0.78 | 1.67 | 97.54 |
|  | 6 | 0.89 | 3.03 | 96.08 |
|  | 7 | 0.91 | 3.00 | 96.08 |
|  | 8 | 0.89 | 3.21 | 95.89 |
|  | 9 | 1.12 | 2.87 | 96 |
| 5 | 00 | 5.15 | 32.96 | 61.88 |
|  | 5 | 1.64 | 0.94 | 97.42 |
|  | 6 | 1.90 | 28.7 | 69.39 |
|  | 7 | 1.97 | 28.95 | 69.07 |
|  | 8 | 2.6 | 25.63 | 71.77 |
|  | 9 | 3.04 | 31.07 | 65.88 |
| 8 | 00 | 7.03 | 46.69 | 46.27 |
|  | 5 | 16.23 | 0.19 | 83.75 |
|  | 6 | 3.99 | 29.62 | 66.39 |
|  | 7 | 3.07 | 30.3 | 66.62 |
|  | 8 | 5.02 | 24.35 | 70.63 |
|  | 9 | 6.87 | 39.64 | 53.48 |
| 10 | 00 | 12.08 | 48.85 | 39.07 |
|  | 5 | 29.66 | 0.17 | 70.16 |
|  | 6 | 5.76 | 26.45 | 67.78 |
|  | 7 | 8.79 | 23.71 | 67.50 |
|  | 8 | 12.55 | 20.76 | 66.68 |
|  | 9 | 13.29 | 38.26 | 48.44 |
| 13 | 00 | 20.04 | 46.30 | 33.66 |
|  | 5 | 47.41 | 0.15 | 52.43 |
|  | 6 | 15.72 | 20.30 | 63.98 |
|  | 7 | 21.96 | 19.66 | 58.38 |
|  | 8 | 25.2 | 15.78 | 59.02 |
|  | 9 | 24.18 | 30.44 | 45.37 |